\documentclass[structabstract]{aa}
\usepackage{natbib}
\usepackage{graphicx}
\usepackage{txfonts}
\def\arcsec{\hbox{$^{\prime\prime}$}}

\begin{document}
   \title{The redshifted network contrast of transition region emission}
   \author{W. Curdt\inst{1}
          \and H. Tian\inst{1, 2}
          \and B. N. Dwivedi\inst{1, 3}
          \and E. Marsch\inst{1}
          }
   \institute{Max-Planck-Institut f\"ur Sonnensystemforschung (MPS),
   Max-Planck-Str.2, 37191 Katlenburg-Lindau, Germany\\
   \email{curdt@mps.mpg.de}
\and School of Earth and Space Sciences, Peking University, China
\and Department of Applied Physics, Institute of Technology, Banaras Hindu University, Varanasi-221005, India
             }
   \date{Received July 1, 2008; accepted October 1, 2008}
  \abstract
   {}
   {We study the VUV emission of the quiet Sun and the net redshift of transition
   region lines in the SUMER spectral range. We aim at establishing a link with atmospheric processes and
   interpreting the observed downflow as the most evident part of the prevailing
   global coronal mass transport.}
   {We rank and arrange all pixels of a monochromatic raster scan by radiance and define
   equally-sized bins of bright, faint, and medium-bright pixels.
   Comparing the bright pixels with the faint pixels, we determine the
   spectrally-resolved network contrast for 19 emission lines. We then
   compare the contrast centroids of these lines
   with the position of the line itself. We establish a relationship between the
   observed redshift of the network contrast with the line formation temperature.}
   {We find that the network contrast is offset in wavelength compared to
   the emission line itself. This offset, if interpreted as redshift, peaks at middle transition
   region temperatures and is 10 times higher than the previously reported net
   redshift of transition region emission lines.
   We demonstrate that the brighter pixels are more redshifted, causing both
   a significant shift of the network contrast profile
   and the well-known net redshift. We show that this effect can be reconstructed
   from the radiance distribution. This result is compatible with loop models,
   which assume downflows near both footpoints.}
   {}
   \maketitle
%
\section{Introduction}
Observations and interpretations of red- and/or blueshifted emission lines
from cosmic objects are crucial to understand the physical processes at work there.
The net redshift in the solar transition region (TR) emission lines has been known
since the Skylab era \citep[e.g.,][and references therein]{Doschek76}.
Redshifts have also been recorded in stellar spectra \citep{Ayres83,Wood96}.
More recently, \citet{Brekke97} and \citet{Chae98} independently verified
this result, analysing high spectral resolution observations from the
Solar Ultraviolet Measurements of Emitted Radiation (SUMER)
instrument on $SoHO$ \citep{Wilhelm95}. Both these groups found similar results for the
quantitative dependence of the net redshift on line formation temperature in the
range from 10$^4$ K to 10$^6$ K. The reported peak downflow of 6 to 8 km/s at a temperature
of around 10$^5$ K is four times higher than the detection limit of the instrument.
However, both groups adopted incorrect literature values for
the rest wavelengths for the Ne\,{\sc viii} and Mg\,{\sc x} emission.  \citet{Dammasch99}
and \citet{Peter99} have established more realistic rest wavelength values
for these species and have demonstrated the disappearance of the net redshift
in coronal emission lines. To our knowledge, a satisfactory physical explanation
of the net redshift has not yet been found. \citet{Peter06} has recently
reported the first full-Sun VUV emission line profile, showing enhanced
emission in the wings.

\begin{figure}[thb]
\centering
\includegraphics[width=9cm]{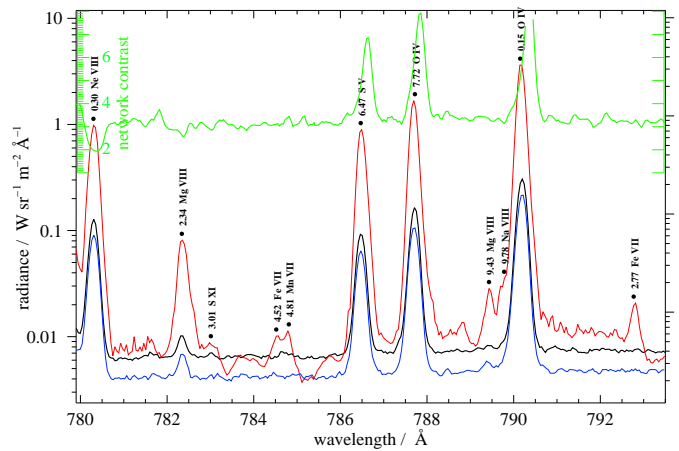}
\caption{Enlarged portion of the SUMER spectral atlas \citep{Curdt01} showing
radiances of average QS (black), sunspot (red), and coronal hole (blue) regions.
The network contrast (ratio bright network\,/\,cell interior in green), which is
normally around 3, increases to values of 6 to 8 in TR lines, and the centroids
of the contrast profiles are clearly redshifted.}
\label{atlas}
\end{figure}
We present a new method to investigate and explain the TR redshift using the
network contrast. A spectrally resolved contrast curve has been
included in the SUMER disk atlas of \citet[hereafter referred to as SDA]{Curdt01}.
Here, the network contrast -- defined as the radiance ratio of pixels in the
bright network and pixels in the cell interior -- has values of about 3 in
the continua, and rises to values of  6 to 8 in TR emission lines. A
similar result was reported by \citet{Reeves}.
In coronal lines such as Ne\,{\sc viii} and Mg\,{\sc x} the contrast is
below the background value, a finding which is equivalent to the result of \citet{Doschek06},
who reported a low correlation between the emission of the TR and corona.

An enlarged cutout of the SDA quiet-Sun profile -- dominated by the strong
Ne\,{\sc viii}, S\,{\sc v}, and O\,{\sc iv} emission
lines around 785 {\AA} -- is displayed in Fig.1.
It is also obvious, although not explicitly mentioned in the SDA, that the
network profiles are redshifted compared to the emission lines themselves.
Our goal is to give a physical explanation for this offset.

In this paper, we extend the earlier work of \citet{Reeves} and the work reported
in the SDA by a comprehensive investigation of the contrast employing a much larger data set.
We show that our result is a direct consequence of the redshift-to-brightness relationship
and can be reconstructed by a simple model using multi-component contributions to the line profile.
A full discussion of the implications for loop models is
beyond the scope of this work. We only present here some salient features,
and a detailed study taking account of atmospheric models will be covered in a separate paper.

In contrast to the earlier work of \citet{Brekke97} and \citet{Chae98}, our new indirect
method is unique in several ways, namely
\newline  (i) it does not require an accurate wavelength calibration,
\newline (ii) it is independent of an exact knowledge of the rest wavelength,
\newline(iii) it closely relies on physical processes in the solar atmosphere.

\section{Method}
The signal in each pixel is a mixture of emission from different plasmas
along the LOS and from unresolved fine structures. We have decomposed them
by a statistical method based on the assumption that statistically
different components can be separated in the radiance distribution.
In our new method, we differentiate among different classes of brightness in
the radiance distribution instead of simply averaging the brightness over all pixels.
We assume that individual bins of different brightness do behave differently. In particular,
we make use of the fact that brighter pixels have a tendency to appear
redshifted in many emission lines. Such a redshift-to-brightness relationship was
also noted by \citet{Dammasch08} in a different data set, and must consequently have an imprint
on the network contrast. This is the core of our method: the strong
redshift observed in the network contrast as compared to the position of the line itself.
For a collection of prominent, blend-free emission lines, we determined the network
contrast (ratio of 33\% of the brighter pixels as compared to 33\% of the dim
pixels) and compared the contrast curve to the emission line itself.

A rather crude and empirical method was used in the SDA to display the network
contrast; the radiance of a few bright pixels out of 300 along the slit,
which were thought to represent the network,
was compared to the radiance of the remaining pixels. The pixel selection was made in
a static way for all 36 individual exposures of the data set, a procedure that may not
be appropriate in view of the temporal evolution of the Sun. However, the
shift of the slit image caused by the wavelength scan
(due to misalignment of the grating) had
been compensated for by employing the standard delta\_pixel routine.

In the present analysis, we measure and compare the position of the line
centroids (the line center determined by spectral centroiding)
relative to the position of the contrast maximum or minimum. We find that the centroid
of the contrast profile is normally shifted by several pixels towards longer wavelengths.

\section{Observations}

The data set in the SDA is a snapshot of 300 pixels
along the slit. Thus, the given spectral radiances still have significant uncertainties
and can only approximately represent the quiet Sun.
Also, better values for the network contrast could have been achieved if a better
statistical basis were available like the one we report here, where rasters
are observed instead of a single exposure.
Our data set consists of raster scans of size 51{\arcsec}~$\times$~120{\arcsec}
in 14 different wavelength windows of $\approx$ 44~{\AA} covering the entire
wavelength range from 670~{\AA} to 1490~{\AA} with only few insignificant
gaps. We employed a slit of size 1{\arcsec}$\times$~120{\arcsec}.
The raster increment was 1.5{\arcsec} and the exposure time was 90~s. The
rasters were obtained in the so-called 'Schmierschritt' mode, which means that each
transmitted spectrum is composed of four elementary exposures with a 22.5~s dwell
time and a 0.375{\arcsec} step increment \citep{Wilhelm95}.
We already compensated on board for the parasitic movement of the slit image
mentioned earlier. Therefore, we can safely assume that the North-South offset between
the individual rasters is negligible. We have compensated for the solar rotation
after each raster scan. Therefore, we also assume that all rasters do map the
same portion of the Sun in the East-West direction.

Our data set was obtained during an observation on 5 April 2007 running from
00:51 UTC to 13:10 UTC. The initial pointing -- centre of the first raster --
was $x=0$, $y=0$. Fifty minutes are needed for each raster. Standard procedures
from the SUMERsoft library\footnote{SUMERsoft is a software library, which constitutes
the integrated experience with SUMER data analysis tools. It is available at
http://www.mps.mpg.de/projects/soho/sumer/text/list\_sumer\_soft.htm}
were applied for the data reduction.

The major improvement of this observation (called the 'super atlas') as
compared to a normal reference spectrum is the increase of the number of
pixels by more than an order of magnitude.
This data set allows us to produce monochromatic raster scans for all emission lines and all
continua in the SUMER spectral range. As an example, we show in Fig. 2 the maps obtained
simultaneously in the emission of N\,{\sc iv}, Ne\,{\sc viii} and in the continuum around 780~{\AA}.

\begin{figure}
   \centering
   \includegraphics[width=8.8cm]{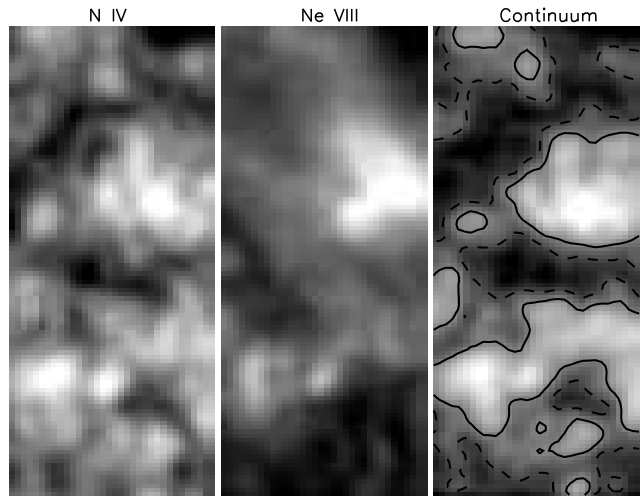}
      \caption{Raster scan in the emission of N\,{\sc iv}, Ne\,{\sc viii} and the continuum around
      780~{\AA}. In the continuum map brightness contours at 33\% and 67\% levels
      are overlaid. These have been used to define bright, faint and
      medium-bright pixels.}
\label{raster}
\end{figure}

\begin{figure*}
\sidecaption
\includegraphics[width=13cm]{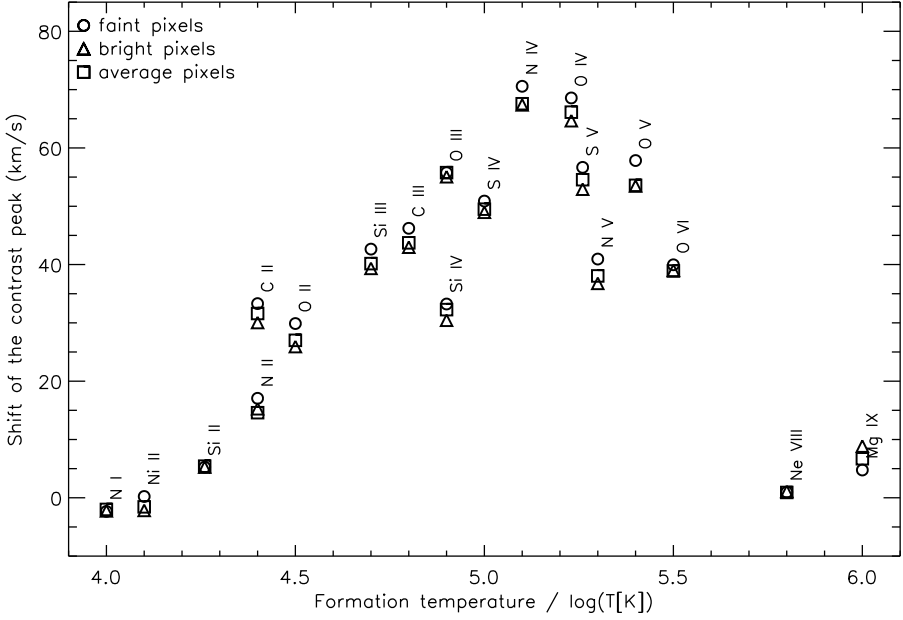}
\caption{The shift of the network contrast relative to the position of the emission line
as a function of formation temperature. Individual offsets have been
determined for comparison with the profile of the bright pixels ($\bigtriangleup$),
the faint pixels ($\circ$), and the average pixels ($\Box$). The shift is scaled
as Doppler flow. This is to make it comparable to earlier, direct measurements and
should not be confused with a real flow.}
\label{contrast}
\end{figure*}

We have selected 19 prominent and blend-free emission lines to produce monochromatic
maps. For each raster we also produce a map of the continuum.
We use this continuum map, where the elements of the chromospheric network
are well-structured and at instrument resolution, to rank all pixels by radiance.
We define equally-sized bins of bright network pixels, of faint cell-interior pixels
and of pixels with medium brightness. The same bin definition was used
for all emission lines in a raster for the determination of the spectrally
resolved contrast profile.
For each spectral pixel the ratio of the radiance found in the bright-pixel
bin over that of the faint pixel bin was determined.
Like in the SDA, a significant increase of the contrast is observed for all lines except for
N\,{\sc i}, Ni\,{\sc ii}, Si\,{\sc ii}, Ne\,{\sc viii}, and Mg\,{\sc ix},
which show a contrast minimum. To account for temporal variations,
we have repeated the definition of radiance bins for each raster.

All selected lines are listed in Table~1 with wavelength and formation temperature and
with the offset results and timing information of the relevant raster.
In order to compare our result with previous work, we have converted the
offsets to Doppler flows, applying $v/c = \Delta \lambda / \lambda$.
Fig.~3 displays the main result of our work.

\begin {table}
\caption{Observed offset of the network contrast for 19 emission lines. Lines
from the same raster are co-temporal, raster start and stop time are in UTC.}
\begin {tabular}{lrccc}
\hline\hline
line & $\lambda$ / {\AA} & log $T$/K & $\Delta \lambda$ / {\AA}   &time\\
\hline
Mg\,{\sc ix}  &749.54    & 6.0 &  0.017 & 01:44 - 02:35\\
S\,{\sc iv}   &750.22    & 5.0 &  0.124 & 01:44 - 02:35\\
N\,{\sc iv}   &765.15    & 5.1 &  0.172 & 02:37 - 03:28\\
Ne\,{\sc viii}&780.30    & 5.8 &  0.002 & 02:37 - 03:28\\
S\,{\sc v}    &786.47    & 5.26&  0.143 & 02:37 - 03:28\\
O\,{\sc iv}   &787.72    & 5.23&  0.174 & 02:37 - 03:28\\
O\,{\sc v}    &761.99    & 5.4 &  0.136 & 02:37 - 03:28\\
O\,{\sc ii}   &833.32    & 4.5 &  0.075 & 03:30 - 04:20\\
O\,{\sc iii}  &833.74    & 4.9 &  0.155 & 03:30 - 04:20\\
C\,{\sc iii}  &977.03    & 4.8 &  0.142 & 05:15 - 06:06\\
C\,{\sc ii}   &1037.00   & 4.4 &  0.109 & 06:08 - 06:59\\
O\,{\sc vi}   &1031.93   & 5.5 &  0.134 & 06:08 - 06:59\\
N\,{\sc ii}   &1083.99   & 4.4 &  0.053 & 07:01 - 07:52\\
Si\,{\sc iii} &1206.51   & 4.7 &  0.161 & 07:54 - 08:45\\
N\,{\sc v}    &1238.82   & 5.3 &  0.157 & 08:48 - 09:39\\
Si\,{\sc ii}  &1309.28   & 4.26&  0.024 & 10:33 - 11:24\\
Ni\,{\sc ii}  &1317.22   & 4.1 &  -0.007& 10:33 - 11:24\\
N\,{\sc i}    &1318.98   & 4.0 &  -0.009& 10:33 - 11:24\\
Si\,{\sc iv}  &1402.77   & 4.9 &  0.151 & 11:26 - 12:17
\end {tabular}
\end {table}

\begin{figure*}
   \sidecaption
   \includegraphics[width=13cm]{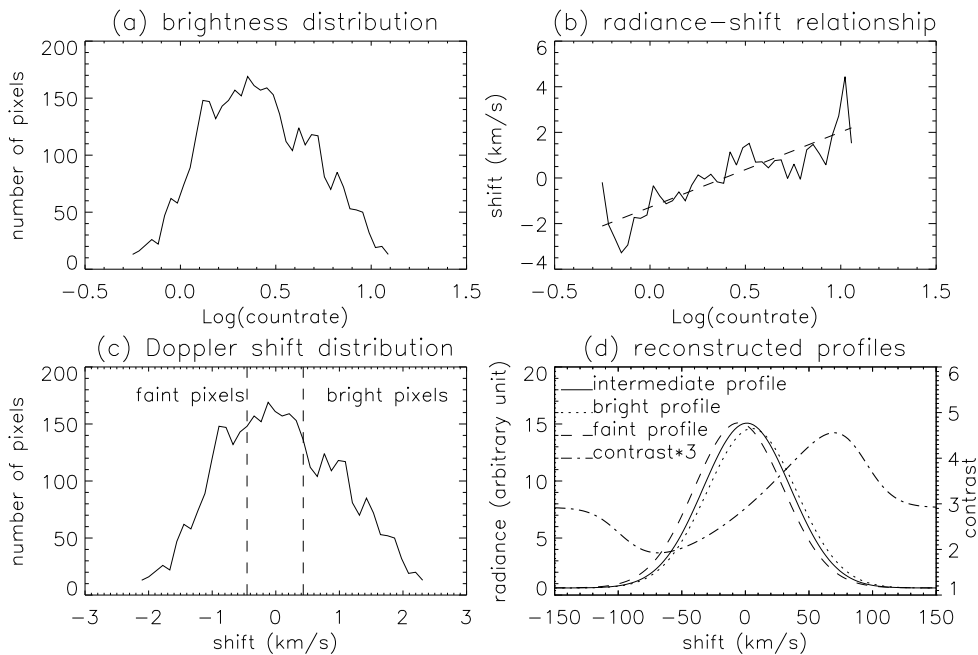}
   \caption{Reconstruction of the observed network contrast in the radiance
   of the $\lambda$765 N\,{\sc iv} TR line~(a) employing the also observed
      redshift-to-brightness relationship~(b) to rescale the abscissa
      of the lognormal radiance distribution~(c)
      and a convolution of three separate groups of pixels with a
      Gaussian profile on a continuum~(d).    }
   \label{model}
\end{figure*}

Individual offsets have been
determined for a comparison with the profile of the bright pixels ($\bigtriangleup$),
the faint pixels ($\circ$), and the average pixels ($\Box$).
In TR emission, the data points for the bright and for the faint pixel profiles
deviate, as can be expected from the earlier work of \citet{Brekke97} and \citet{Chae98}.
This interesting secondary result is a direct consequence of the redshift-to-brightness relationship.

\section{Discussion and summary}

Fig. 3 has similarities to the respective figures in
\citet{Brekke97} and \citet{Chae98}.
We have expressed the offset $\Delta \lambda$ in speed units of Doppler flow
and arrived at values of about 60~km/s. This is about 10 times higher than
the average net downflow found in earlier work. This finding demonstrates
that our new method enhances the visibility of the Doppler shift, which leads
to a significant increase of the sensitivity, but should not be confused with a real flow.

We now use a simple model to reconstruct the observed redshift
of the network contrast. We use the fact that the radiance distribution of
the average quiet Sun follows a single lognormal distribution function
and that a redshift-to-brightness relationship exists.
\citet{Pauluhn00} have shown in great detail that the emission
of the quiet Sun including the network and the intranetwork is better
described by a lognormal distribution than by two Gaussians. This is in
agreement with the brightness distribution of the $\lambda$765 N\,{\sc iv}
TR line, as shown in Fig.4a.

For this distribution, we have defined 40 equally sized
radiance bins. For each bin we determined the line position.
We found a linear relationship between the logarithm of the spectral
radiance, $L_{\lambda}$, and the redshift of the line centroid, as displayed in
Fig.4b. \citet{Dammasch08} applied a different bin definition, but they
arrive at a similar result of 3 km/s per radiance decade. We now use the
redshift-to-brightness relationship to rescale the abscissa in the histogram
in Fig.~4c and define separate groups of bright pixels, of faint pixels and the pixels of
medium brightness. The pixels in each group are convolved with a
Gaussian line profile typical for TR emission on a continuum
background. Thus we arrive at three different profiles for those groups of
pixels. As a consequence of the redshift-to-brightness relationship shown in Fig.4b,
the reconstructed profiles are offset from each other
by a few km/s (cf., Fig.~4d). The contrast profile -- the ratio of the bright
profile to the faint profile -- peaks, however, near a redshift value of
$\approx$70~km/s, which is close to our empirical result in Fig.~3. This demonstrates
that the observed offset of the network contrast can be reconstructed in a
quantitative manner with some basic assumptions. We note that the contrast profile, which
we use in our new method, is much more sensitive to shifts than the line profile itself.
We also note that the observed skewness of the contrast profile in lines like
$\lambda$780 ~Ne\,{\sc viii} or $\lambda$1032~O\,{\sc vi} seems to be a real result
of the reconstruction (cf., Fig.~4d).

We emphasize that our results cannot be used to make any statement about
systematic flows in coronal emission. The contrast depression simply indicates
that the corona is decoupled from the chromosphere, and therefore no deviation
from the rest wavelength can be expected for the contrast minimum.
Cool plasma with a low degree of ionization will not be guided by the magnetic
field and will also not participate in the concentration process in the downflow near loop
footpoints.
The redshift-to-brightness relationship as a direct consequence of this most
evident part of the prevailing global coronal mass transport
\citep{Marsch08}
may be the physical explanation for the well-known net redshift
in TR emission and of the offset in the spectrally resolved network contrast.

Our new result corroborates the recent work of \citet{Dammasch08} and \citet{Marsch08}.
In actuality, blueshifts and redshifts respectively correspond to upflows
and downflows of the plasma on open and closed field lines as noted by
Marsch et al. (2008). The redshifted network contrast of TR emission,
ubiquitous redshifts and sporadic blueshifts in the solar atmosphere show the
physical characteristics of mass transport which we may term as 'coronal convection'.

\citet{Dammasch08} argue that the downflow concentrated
near both footpoints of coronal loops is powered by a quasi-continuous heating process.
They follow the suggestion of \citet{Feldman01} and infer that unresolved bright features
in the network are tiny loops, which are also redshifted at both legs.
\citet{Pauluhn07} developed a heuristic nanoflare model showing that continuous
small-scale brightenings could produce the observed radiance distribution.
\citet{Fontenla07,Fontenla08} assume that the Farley-Bunemann instability could
be responsible for the heating process. It is difficult to use our observation
in favour of or against any nanoflare heating model reported in the literature
\citep[see the review of][and references cited therein]{Innes04}.
These questions can certainly not be answered from our observation alone.
More theoretical work is needed, which is beyond the scope of this
communication.

\begin{acknowledgements}
The SUMER project is financially supported by DLR, CNES, NASA, and the ESA PRODEX
Programme (Swiss contribution). SUMER is part of {\it SOHO} of ESA and NASA.
The work of BND was supported by a grant from the MPS. HT is supported by
China Scholarship Council for his stay at MPS. We thank the referee for critical
comments which improved the clarity of this Letter.
\end{acknowledgements}

\end{document}